\documentclass[preprint,superscriptaddress,showpacs,amsmath,amssymb,aps,prb]{revtex4-1}

\usepackage{graphicx}
\usepackage{hyperref}

\newcommand{\vect}[1]{\ensuremath{\boldsymbol{#1}}}
\newcommand{\lapl}{\ensuremath{\nabla^2}}
\newcommand{\grad}{\ensuremath{\vect{\nabla}}}
\newcommand{\divg}{\ensuremath{\vect{\nabla\cdot}}}
\newcommand{\curl}{\ensuremath{\vect{\nabla\times}}}

\newcommand{\h}{\vect{h}}
\newcommand{\vecr}{\vect{r}}
\newcommand{\rp}{\vect{r'}}
\newcommand{\ave}[1]{\ensuremath{\langle #1 \rangle}}
\newcommand{\vk}{\vect{k}}
\newcommand{\z}{\vect{z}}

\newcommand{\LE}{\ensuremath{\lambda_{\mathrm{eff}}}}
\newcommand{\erfc}{\mathrm{erfc}}
\newcommand{\rz}{\ensuremath{R(0)}}
\newcommand{\rarg}{\ensuremath{R(z)}}
\newcommand{\omegy}{\left(2 \sqrt{\nu} \frac{z}{l} \right )}

\begin{document}

\title{Calculation of the effect of random superfluid density on the  temperature dependence of the penetration depth.}  

\author{Thomas  M. Lippman}
\affiliation{Stanford Institute for Materials and Energy Sciences,
  SLAC National Accelerator Laboratory, 2575 Sand Hill Road, Menlo
  Park, CA 94025, USA.}
\affiliation{Department of Physics, Stanford University, Stanford,
  California 94305-4045, USA.}

\author{Kathryn A. Moler}
\email{kmoler@stanford.edu}
\affiliation{Stanford Institute for Materials and Energy Sciences,
  SLAC National Accelerator Laboratory, 2575 Sand Hill Road, Menlo
  Park, CA 94025, USA.}
\affiliation{Department of Physics, Stanford University, Stanford,
  California 94305-4045, USA.}
\affiliation{Department of Applied Physics, Stanford University,
  Stanford, California 94305-4045, USA.}

\date{\today}

\begin{abstract}
Microscopic variations in composition or structure can lead to
nanoscale inhomogeneity in superconducting properties such as the
magnetic penetration depth, but measurements of these properties are
usually made on longer length scales.  We solve a generalized London
equation with a non-uniform penetration depth, $\lambda(\vecr)$,
obtaining an approximate solution for the disorder-averaged Meissner
screening.  We find that the effective penetration depth is different
from the average penetration depth and is sensitive to the details of
the disorder.  These results indicate the need for caution when
interpreting measurements of the penetration depth and its temperature
dependence in systems which may be inhomogeneous.

\end{abstract}


\maketitle

\section{Introduction}
The penetration depth and its temperature dependence are important
characteristics of any superconductor and are considered key to
determining the momentum space structure of the order parameter.
\cite{hardy_precision_1993,bok_magnetic_2002,prozorov_magnetic_2006}
The possibility of disorder in exotic superconductors is well known,
but analyses performed to date have concentrated on the effect
of disorder-induced scattering on the momentum space structure
of the gap.
\cite{annett_interpretation_1991,mishra_lifting_2009,hirschfeld_effect_1993,vorontsov_superfluid_2009}
This paper is motivated by the possibility that disorder may lead to
nanoscale real space variation and the associated need to model the
relationship between such spatial variation and properties that are
measured on longer length scales.  We address how inhomogeneity in the
penetration depth may affect bulk measurements of the penetration
depth for methods that rely on Meissner screening and can be analyzed
by solutions to London's equation. In particular, we show that the
measured result is not simply given by the average value of the
penetration depth, but is affected by the statistical structure of the
spatial variations in the penetration depth.

Many superconductors are created by chemical doping of a
non-superconducting parent compound. In these systems the inherent
randomness of the doping process may give rise to an inhomogeneous
superconducting state. The importance of this effect will be
determined by the characteristic length over which the dopant atoms
affect the superconductivity. Even in the most ordered material, there
will be binomial fluctuations in the total number of dopants in a
given region. In general, one does not expect significant spatial
variation in materials that are weakly correlated and can be described
by a rigid band model. For example, disorder is largely irrelevant in
classic metallic superconductors, due to their long coherence lengths
and weakly correlated nature.\cite{gennes_superconductivity_1999} In
contrast, the cuprates are doped insulators with a coherence length on
the scale of the lattice. They are known to have nanoscale disorder in
their superconducting properties, as seen by scanning tunneling
microscopy.\cite{fischer_scanning_2007} Similar gap maps have been
observed in the iron pnictide family
\cite{yin_scanning_2009,teague_measurement_2011,fasano_local_2010} and
in disordered titanium nitride films close to the superconductor to
insulator transition.
\cite{sacepe_localization_2011,sacepe_disorder-induced_2008}

Materials with intrinsic disorder present two separate
challenges. Understanding how the random doping process gives rise to
local superconducting properties, such as the penetration depth or
local density of states, requires a microscopic model. But even with
such a model, we still need to make the connection between the local
superconducting properties and bulk measurements. The manner in which
local superconducting properties relate to the observed properties
will differ from experiment to experiment. For instance, a measurement
of the heat capacity will return the total heat capacity of the
macroscopic sample, so the inferred specific heat capacity will be a
volume average over the sample. In contrast, we might expect the
thermal conductivity response to be dominated
by a percolation path connecting regions with small local gap,
$\Delta(\vecr)$, or large local density of states.

Here we focus on the penetration depth, $\lambda$, as measured by
screening of the magnetic field, including both resonant cavity
frequency shift measurements at radio
frequencies\cite{prozorov_magnetic_2006} and the local probes of
Magnetic Force Microscopy\cite{luan_local_2011} and Scanning SQUID
Susceptometry.\cite{hicks_evidence_2009} These methods measure
$\lambda(T)$ by detecting the response magnetic field generated by the
superconductor due to an applied field, and can be analyzed using the
London equation. Thus, we can model the effect of inhomogeneity by
solving the London equation with $\lambda(\vecr)$ as a random function
of position \vecr. Our goal is to find a new equation for the
disorder-averaged magnetic field, as this will determine the measured
response. Here, we work in the limit of small fluctuations to find an
approximate equation for the disorder-averaged magnetic field, as this
will determine the measured response.

\section{Stochastic London Equation}
\label{sec:stoch-lond-equat}
To understand the measured penetration depth when $\lambda(\vecr)$ is
a random function of position, we calculate the disorder-averaged
magnetic field response to obtain an effective penetration depth. For
isotropic and local superconductors in three dimensions, the static
magnetic field $h(\vecr)$ is given by the London equation with
$\lambda(\vecr)$ a function of position.  The correct
form\cite{cave_critical_1986} of the London equation when the
penetration depth is non-uniform is:
\begin{equation}
  \label{london_def}  
  \h + \curl\left[\lambda(\vecr)^2\curl\h\right] = 0,
\end{equation}
which is derived from the second Ginzburg-Landau equation in the
London limit.\cite{london_footnote} We parametrize the penetration
depth as the average value plus a fluctuating term:
\begin{equation}
  \label{lambda_def}
  \lambda(\vecr) = \Lambda\left[1+\xi(\vecr)\right],
\end{equation}
so that $\ave{\lambda(\vecr)} = \Lambda$. Then Eq. \ref{london_def}
becomes:
\begin{equation}
  \label{london_parts}
  \left( L + M_1 + M_2 \right)\h = 0,
\end{equation}
where:
\begin{eqnarray}
  \label{london_part_defs}
  L & \equiv & 1 - \Lambda^2 \lapl \vect{I}, \nonumber \\ 
  M_1 & \equiv & -2 \Lambda^2 \xi \lapl \vect{I} + 2 \Lambda^2
  \grad\xi \vect{\times} \curl, \quad \text{and} \\
  M_2 & \equiv & - \Lambda^2 \xi^2 \lapl \vect{I} + \Lambda^2
  \grad\xi^2 \vect{\times} \curl. \nonumber
\end{eqnarray}
Here \vect{I} is the identity tensor, and the ``dangling curl'' is
understood to operate on a vector to its right. The terms are grouped
so that $M_1$ is first-order in $\xi$, $M_2$ is second-order in $\xi$,
and $L$ gives the unperturbed London equation. We will work in the
limit of small fluctuations, $\xi(\vecr) \ll 1$, so that $ M_1 + M_2$
is a perturbative term in Eq. \ref{london_parts}.

Our method of solution comes from the similarity of the Helmholtz and
London equations. The Helmholtz equation, which governs wave
propagation, becomes the London equation when the wavevector is purely
imaginary. Thus our problem is related to the propagation of waves in
a random medium, and we can build upon a large and multidisciplinary
literature devoted to this
challenge.\cite{mysak_wave_1978,van_kampen_stochastic_1976} The paper by Karal
and Keller \cite{karal_elastic_1964} is particularly relevant, because
it retains the vectorial nature of the problem, rather than
simplifying to a scalar wave equation.

We now derive, from Eq. \ref{london_parts}, an approximate equation
for the disorder-averaged field \ave{\h}. Applying the inverse of $L$
to both sides:
\begin{equation}
  \left[  1 + L^{-1} ( M_1 + M_2 )\right]\h = \h_0,
\end{equation}
where $L\, \h_0 = 0$. Solving for \h:
\begin{equation}
  \h = \left[ 1 + L^{-1} (M_1 + M_2) \right ]^{-1} \h_0,
\end{equation}
assuming the inverse exists. Averaging both sides:
\begin{equation}
  \ave{\h} =   \left\langle \left[ 1 + L^{-1} (M_1 + M_2) \right]^{-1} \right
  \rangle \h_0,
\end{equation}
where $\h_0$ comes outside of the average because it is non-random. Solving
for $\h_0$:
\begin{equation}
  \left\langle \left[ 1 + L^{-1} (M_1 + M_2) \right]^{-1} \right
  \rangle ^{-1} \ave{\h} = \h_0.
\end{equation}
Since we assume small fluctuations, we can expand the term inside the
average:
\begin{equation}
  \left\langle 1 - L^{-1} (M_1 + M_2) + L^{-1} M_1 L^{-1} M_1 +
  \mathcal{O}(\xi^3) \right \rangle ^{-1} \ave{\h} = \h_0.
\end{equation}
Averaging and expanding again:
\begin{equation}
  \left(  1 - L^{-1} \ave{M_1 L^{-1} M_1} + L^{-1}\ave{M_2} \right
  )\ave{\h} = \h_0,
\end{equation}
since $\ave{M_1} = 0$ due to Eq. \ref{lambda_def}. We then apply $L$
to both sides:
\begin{equation}
  \label{ave_eqn}
  \left( L - \ave{M_1 L^{-1} M_1} + \ave{M_2} \right )\ave{\h} = 0,
\end{equation}
which yields the average field to second order in $\xi$.

\section{Results}
We first evaluate the averages in Eq. \ref{ave_eqn}, giving us an
equation for \ave{\h}\ in terms of the penetration depth correlation
function, \ave{\lambda(\vecr)\lambda(\rp)}. We then consider two
specific cases for the correlation function and numerically evaluate
the effective penetration depth for a range of parameters.

\subsection{Evaluating the Averages}
\label{sec:evaluating-averages}
We will solve Eq. \ref{ave_eqn} for a single Fourier mode of $
\ave{\h(\vecr)} = \h\, e^{i \vk\cdot\vecr}$, then derive an equation
for \vk\ that yields exponentially decaying solutions consistent with
Meissner screening.

First, we evaluate \ave{M_2}:
\begin{equation}
  \ave{M_2} = - \Lambda^2\, \ave{\xi(\vecr)^2}\, \lapl I + \Lambda^2\,
  \ave{\grad\xi(\vecr)^2\,} \vect{\times} \curl.
\end{equation}
We introduce the correlation function $R(\vecr,\rp) =
\ave{\xi(\vecr)\xi(\rp)}$, which is a function only of $|\vecr - \rp|$
if $\xi$ is stationary and isotropic. Then we see that $\ave{\xi^2} =
\rz$ and $\ave{\grad\xi^2} = \grad \ave{\xi^2} = 0$, so
\begin{equation}
  \label{ave_M2}
  \ave{M_2}\ave{\h} =  \Lambda^2\, k^2 \,\rz\, \h\,
  e^{i\vk\cdot\vecr}.
\end{equation}

We now evaluate the remaining average, $\bigl \langle M_1 L^{-1}
M_1\bigr \rangle$, in three stages to derive
Eq. \ref{ave_integrated_M1_L_M1}.  First we expand the differential
operations, then evaluate the disorder average. The last stage is to
evaluate the integral. We will then combine this integral with
Eq. \ref{ave_M2} to solve Eq. \ref{ave_eqn}.

The average to evaluate has the form:
\begin{equation}
  \label{explicit_form_M1}
  \bigl\langle M_1 L^{-1} M_1\bigr\rangle \ave{\h} = \int d\rp \Bigl\langle
  M_1(\vecr)G(\vecr-\rp)M_1(\rp)\Bigr\rangle \ave{\h(\rp)}.
\end{equation}
The Green's function is the solution to   $ (1-\Lambda^2\lapl)
\,G(\vecr,\rp) = \delta(\vecr-\rp)$, and is:
\begin{equation}
  \label{Greens_Fn}
  G(z) = \frac{1}{\Lambda^2}\frac{1}{4\pi z}\,e^{-z/\Lambda}.
\end{equation}
Here, we have defined $\z = \vecr - \rp$.

We now expand the differential operations in
Eq. \ref{explicit_form_M1}. We do this in two segments, first with
derivatives at \vecr, then with derivatives at \rp. The first is:
\begin{multline}
  \label{first_half_M1}
  M_1(\vecr)G(\vecr-\rp)\vect{v}(\rp) = \bigl[-2
    \Lambda^2\xi(\vecr)\lapl_{\vecr}+ 2
    \Lambda^2\grad_{\vecr}\xi(\vecr)\times\nabla_{\vecr}\times
    \bigr]G(\vecr-\rp)\vect{v}(\rp)\\ = 2 \xi
  \vect{v}(\rp)\bigl[\delta(\z)-G(\z)\bigr] + 2 \Lambda^2
  \grad\xi\times\bigl[\nabla_{\vecr}G(\vecr-\rp)\bigr]\times\vect{v}(\rp).
\end{multline}
 The second, which was represented by \vect{v(\rp)} above, is:
\begin{multline}
  \label{second_half_M1}
  \vect{v}(\rp) = M_1(\rp)\ave{\h(\rp)} =
  e^{i\vk\cdot\rp}\Bigl[2\Lambda^2 k^2\xi(\rp)\h + 2 i \Lambda^2
    \grad_{\rp}\xi(\rp)\times\vk\times\h\Bigr]\\ = 2 \Lambda^2
  e^{i\vk\cdot\rp}
  \Bigl\lbrace\bigl[k^2\xi(\rp)-i\,\grad_{\rp}\xi(\rp)\cdot\vk\bigr]\h + i
    \bigl[\grad_{\rp}\xi(\rp)\cdot\h\bigr]\vk\Bigr\rbrace.
\end{multline}
Combining Eqns. \ref{first_half_M1} and \ref{second_half_M1}, we obtain:
\begin{multline}
  \label{M1_L_M1}
  M_1(\vecr)G(\vecr-\rp)M_1(\rp) \ave{\h(\rp)} = 2 \Lambda^2
  e^{i\vk\cdot\rp} \Biggl\lbrace \xi(\vecr)\xi(\rp) \bigl[\delta(z) -
  G(z)\bigr] 2 k^2 \h \\
  + \xi(\vecr)\grad_{\rp}\xi \bigl( \h \otimes \vk - \vk \otimes \h
    \bigr) 2 i \bigl[ \delta(z) - G(z) \bigr]\\
  -2 \Lambda^2 G(z) \bigl(\Lambda^{-1} + z^{-1} \bigr) \biggl [
  \xi(\rp)\grad_{\vecr}\xi \bigl( \h \otimes \hat{\z} - \hat{\z}
    \otimes \h \bigr) k^2 \\
  + i \bigl( \h \otimes \hat{\z} - \hat{\z} \otimes \h \bigr)
  \bigl (\grad_{\vecr}\xi \otimes \grad_{\rp}\xi \bigr) \vk 
  + i \bigl( \hat{\z} \otimes \vk - \vk \otimes \hat{\z} \bigr) \bigl
  (\grad_{\vecr}\xi \otimes \grad_{\rp}\xi \bigr) \h \biggr ]
  \Biggr \rbrace,
\end{multline}
where we use $\otimes$ to indicate the tensor product.

To perform the disorder average in the second stage, we need various
derivatives of the correlation function \rarg:
\begin{equation*}
  \begin{split}
    \bigl\langle \xi(\vecr)\,\grad_{\rp}\xi \bigr\rangle &=
    \grad_{\rp} \,R\bigl(|\vecr - \rp | \bigr)  =
    -\hat{\z}\,\dot{R}(z),\\
    \bigl\langle \xi(\rp)\,\grad_{\vecr}\xi \bigr\rangle &=
    \grad_{\vecr} \,R\bigl(|\vecr - \rp | \bigr)  =
    \hat{\z}\,\dot{R}(z), \quad \text{and}\\ 
    \bigl\langle \grad_{\vecr}\xi\,\grad_{\rp} \xi \bigr\rangle &=
    \grad\grad'\,R(|\vecr-\rp|) =  -
    \biggl[ \frac{\dot{R}}{z} \vect{I}
      +\hat{\z} \otimes \hat{\z}\,\Bigl (\ddot{R}-
      \frac{\dot{R}}{z} \Bigr)\biggr],
  \end{split}
\end{equation*}
where the overdot indicates differentiation with respect to z, and
\vect{I}\ is the identity tensor. Then averaging
Eq. \ref{M1_L_M1} gives: 
\begin{multline}
  \label{ave_M1_L_M1}
  \bigl\langle M_1(\vecr)G(\vecr-\rp)M_1(\rp)\bigr \rangle
  \ave{\h(\rp)} = 2 \Lambda^2 e^{i\vk\cdot\rp} \Bigl \lbrace \bigl[
    A(z) + 2 k^2 \rarg \delta(z) \bigr] \h \\
  - \bigl [ B(z) + 2 i \dot{R}(z) \delta(z) \bigr] \bigl( \vk
    \otimes \h - \h \otimes \vk \bigr) \hat{\z}
  - C(z) \h \bigl( \hat{\z} \otimes \hat{\z} \bigr)\Bigr \rbrace,
\end{multline}
with the scalars $A, B$, and $C$ given by:
\begin{eqnarray} 
  A(z) &=& \ 2 k^2 \Bigl[ \dot{R}(z) \Lambda^2 G(z)
    \bigl(\Lambda^{-1} + z^{-1} \bigr) - \rarg \, G(z) \Bigr],
  \nonumber \\
  B(z) &=& \ 2 i \Bigl[ \ddot{R}(z) \Lambda^2
    G(z)\bigl(\Lambda^{-1} + z^{-1} \bigr) - \dot{R}(z)\, G(z)
    \Bigr], \quad \text{and} \nonumber \\
  C(z) &=& \ 2 \Lambda^2 k^2 \dot{R} (z)\, G(z) \bigl(\Lambda^{-1} +
  z^{-1} \bigr). \nonumber 
\end{eqnarray}

The final stage in evaluating Eq. \ref{explicit_form_M1} is to perform
the integral over \rp. We first change variables from \rp\, to \z,
then integrate over the orientation of \z. Using the relations
\begin{equation}
  \begin{split}
    \int\! \mathrm{d}\hat{\z}\ e^{-i\vk\cdot\z} &= 4\pi
    \frac{\sin(kz)}{kz} \equiv F(k,z),\\ \int\!
    \mathrm{d}\hat{\z}\ \hat{\z}\, e^{-i\vk\cdot\z} &= \hat{\vk}\,
    \frac{i}{z}\, \partial_k F, \quad \text{and} \\ \int\!
    \mathrm{d}\hat{\z}\ \hat{\z} \otimes \hat{\z}\, e^{-i\vk\cdot\z}
    &= \frac{-1}{z^2} \left[ \frac{\partial_k F}{k} \vect{I} +
    \hat{\vk} \otimes \hat{\vk} \left(\partial^2_k F -
    \frac{\partial_k F}{k}\right) \right],
  \end{split}
\end{equation}
we find that Eq. \ref{explicit_form_M1} evaluates to:
\begin{equation}
  \label{ave_integrated_M1_L_M1}
  \int d\rp \Bigl\langle M_1(\vecr)G(\vecr-\rp)M_1(\rp)\Bigr\rangle
  \ave{\h(\rp)} = 4 \Lambda^2 k^2 e^{i \vk \cdot \vecr} \Bigl\lbrace
    \left[X + \rz\right]\, \h + Y\, \hat{\vk} (\h \cdot
    \hat{\vk} ) \Bigr\rbrace.
\end{equation}
The functions $X$ and $Y$ are given by:
\begin{multline}
  \label{ave_integ_M1_L_M1_parts}
  X(k) = \int_0^\infty \! \mathrm{d}z\ G(z) \biggl\lbrace
    \Lambda^2 \bigl(\Lambda^{-1} + z^{-1} \bigr) \Bigl[ \dot{R}
      \bigl( z^2 F + k^{-1} \partial_k F \bigr) - \ddot{R} z
      k^{-1} \partial_k F \Bigr] + \dot{R} z k^{-1} \partial_k F
    - R z^2 F \biggr\rbrace, \\
  Y(k) = \int_0^\infty \! \mathrm{d}z\ G(z) \biggl\lbrace \Lambda^2
    \bigl(\Lambda^{-1} + z^{-1} \bigr) \Bigl[ \ddot{R} z k^{-1}
      \partial_k F + \dot{R} \bigl( \partial^2_k F - k^{-1}
      \partial_k F \bigr) \Bigr] - \dot{R} z k^{-1} \partial_k F
    \biggr \rbrace.
\end{multline}
We require the average magnetic field to have $\divg \h = 0$, which
means that $ \vk \cdot \h = 0$. We now collect our results from
Eqns. \ref{ave_integrated_M1_L_M1} and \ref{ave_M2}, and insert them
into Eq. \ref{ave_eqn}:
\begin{equation}
  \label{eval_ave_eqn}
  \h\, e^{i \vk \cdot \vecr} \, \Bigl[ 1 + \Lambda^2 k^2 \, \bigl( 1 -
  3\, \rz - 4\, X \bigr) \Bigr] = 0.
\end{equation}
We are interested in solutions consistent with Meissner screening, so
we require that $k$ be positive and purely imaginary. Then the field
decays on a length scale $ \LE = \tfrac{i}{k},$ which we identify as
the experimentally measured penetration depth. To calculate \LE, we
will solve the equation:
\begin{equation}
  \label{eval_ave_eqn_LAM}
  \frac{\LE^2}{\Lambda^2} = 1 - 3 \rz - 4 X.
\end{equation}
Inserting $k = \tfrac{i}{\LE}$ into our equation for $X$, we get:
\begin{multline}
  \label{X_gped_sincos}
  X = 4 \pi \int_0^\infty \! \mathrm{d}z\ G(z)
  \mathrm{sinh}\negmedspace\left(\frac{z}{\LE} \right) \biggl \lbrace \Lambda^2
  \bigl(\Lambda^{-1} + z^{-1} \bigr) \Bigl[ \dot{R} \LE\ z^{-1}
    \bigl(z^2 + \LE^2 \bigr) - \ddot{R} \LE^3 \Bigr] +
  \dot{R}\LE^3 - R z \LE \biggr \rbrace \\
  + 4 \pi \int_0^\infty \! \mathrm{d}z\ G(z)
  \mathrm{cosh}\negmedspace\left(\frac{z}{\LE} \right ) \biggl \lbrace \Lambda^2
  \bigl(\Lambda^{-1} + z^{-1} \bigr) \Bigl[ - \dot{R}\LE^2 +
    \ddot{R} z \LE^2 \Bigr] - \dot{R} z \LE^2 \biggr
  \rbrace.
\end{multline}
Valid solutions for \LE\ will require the integral for $X$ to converge
and Eq. \ref{eval_ave_eqn_LAM} to have solutions.

\subsection{Correlation Function}
A full solution of the disorder-averaged magnetic field, \ave{\h},
requires knowledge of the correlation function \rarg\ and hence
requires not only a detailed knowledge of the composition, structure,
and disorder of the sample, but also a microscopic model to locally
determine the superconducting properties from that structure. Without
guidance from microscopic calculations, we will use the Mat\'{e}rn
one-parameter family of correlation
functions\cite{matern_spatial_1986} to tune the smoothness, as well as
the magnitude and correlation length, of the penetration depth
fluctuations. Handcock and Wallis \cite{handcock_approach_1994}
parametrize the Mat\'{e}rn class of covariance functions as:
\begin{equation}
  \label{matern}
  \rarg = \frac{\rz}{2^{\nu-1} \Gamma(\nu)}\, \left( 2
  \sqrt{\nu}\, \frac{z}{l} \right)^\nu K _\nu \left(2
  \sqrt{\nu}\, \frac{z}{l} \right ),
\end{equation}
where $K_\nu$ is a modified Bessel function of the second kind and
$\Gamma(z)$ is the Gamma function.  The intercept at zero separation
is the normalized variance of the penetration depth, $\rz =
\sigma_\lambda^2/\ave{\lambda}^2 = (\ave{\lambda^2} -
\ave{\lambda}^2)/\ave{\lambda}^2$, and quantifies the magnitude of the
inhomogeneity in $\lambda(\vecr)$. The correlation length, $l$,
controls the size of the fluctuations in $\lambda(\vecr)$. The
parameter $\nu$ controls the smoothness of $\lambda(\vecr)$. Larger
$\nu$ gives a smoother random field, since it is $\lceil \nu \rceil -
1$ times mean squared differentiable, where $\lceil \cdot \rceil$ is
the ceiling function.\cite{handcock_approach_1994}

Two members of the family deserve specific mention. When $\nu = 1/2$,
Eq. \ref{matern} reduces to the exponential correlation function,
$\rarg = \rz\, \mathrm{exp}\! \left(-z \sqrt{2}/l \right) $, which is
the correlation function of a Markov process in one dimension. The
integrals for $X$ in Eq. \ref{X_gped_sincos} diverge when $\rarg
\propto e^{-z}$, making the case $\nu = 1/2$ invalid. In the limit
where $ \nu \rightarrow \infty$, $\rarg \rightarrow \rz\,
\mathrm{exp}\!  \left(-z^2/l^2\right ) $, which is labeled the squared
exponential correlation function, to prevent confusion with the
Gaussian probability distribution. This correlation function gives the
smoothest possible $\lambda(\vecr)$ that can be described within the
Mat\'{e}rn covariance family.

\subsection{Squared Exponential Correlations}
\label{sec:squar-expon-corr}
\begin{figure}[p]
  \includegraphics[width = 0.8\columnwidth]{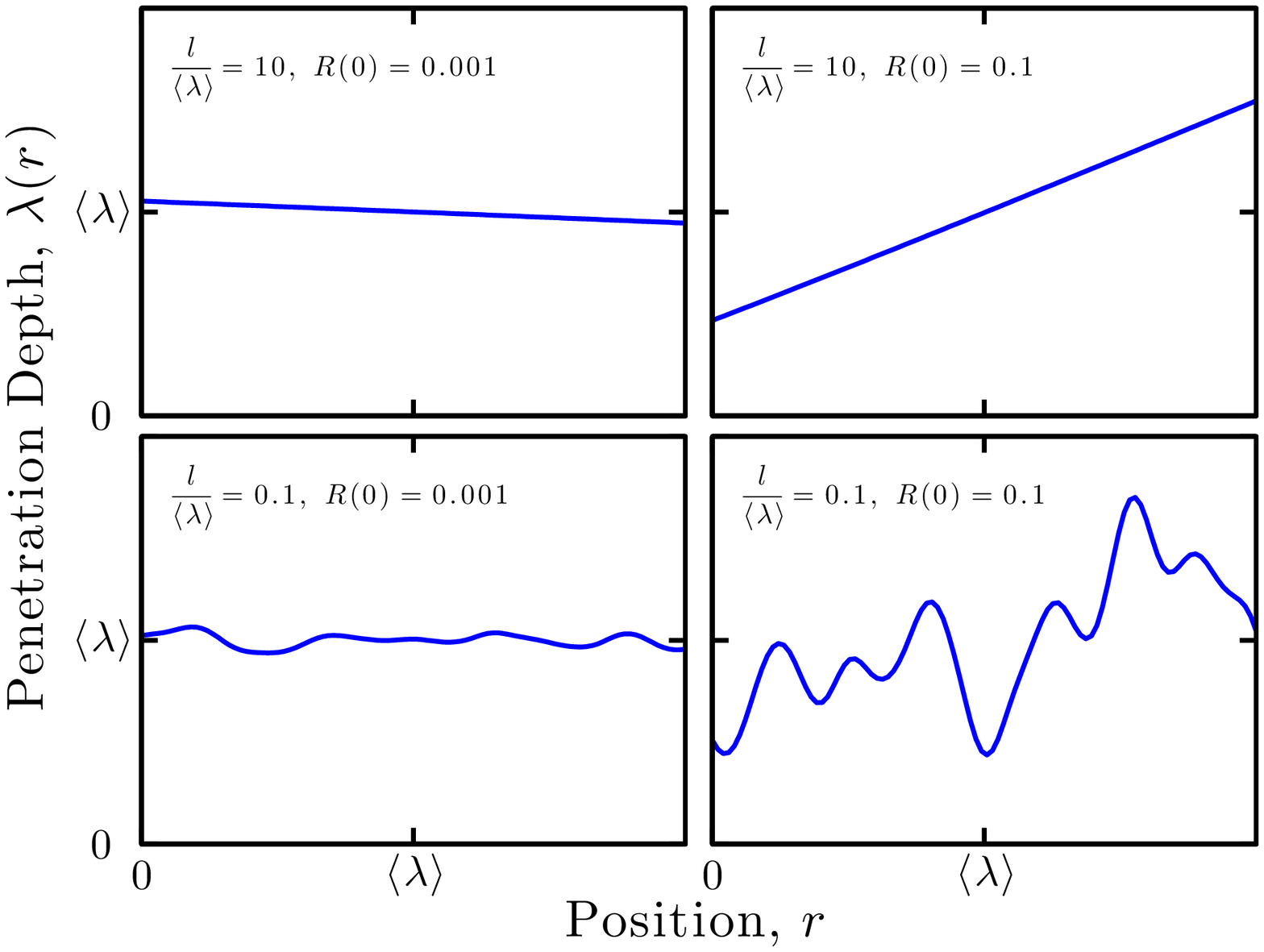}
  \caption{\label{randLam} Sample realizations of a random penetration
    depth reveal the influence of the correlation length, $l$, and
    variance, \rz. The variance, $\rz = \ave{ \lambda^2}/
    \ave{\lambda}^2 - 1 $, controls the width the penetration depth
    distribution, and the correlation length establishes the
    characteristic length over which $\lambda(\vecr)$ changes.}
\end{figure}

We now consider the case of squared exponential correlations, $\rarg =
\rz\,e^{-z^2/ l^2}$.  In Fig. \ref{randLam} we plot four realizations
of a normally distributed penetration depth with squared exponential
correlations, illustrating the effect of the two
parameters $l$ and $\rz$ on $\lambda(\vecr)$. Evaluating
Eq. \ref{X_gped_sincos} gives:
\begin{multline}
  \label{X_insert_gauss_Gamma}
  X = -\rz \int_0^\infty \! \mathrm{d}z\ e^{-z/\Lambda}
  e^{-z^2/l^2} \mathrm{sinh}\negmedspace\left(\frac{z}{\LE} \right
  )\ \frac{\LE}{\Lambda^2}\, \Bigl [  \bigl ( 1 + 2
    \tfrac{\Lambda^2}{l^2} \bigr) + 2\,z\, \frac{\Lambda}{l^2}
    \Bigr]\ \big(1 + 2 \tfrac{\LE^2}{l^2} \bigr) \\
  + 2\,\rz \int_0^\infty \! \mathrm{d}z\ e^{-z/\Lambda} e^{-z^2/2
    l^2} \mathrm{cosh}\negmedspace\left(\frac{z}{\LE} \right )
  \frac{\LE^2}{\Lambda^2} \Bigl [ z\, \frac{1}{l^2} \bigl(1 + 2\,
    \tfrac{\Lambda^2}{l^2} \bigr) + 2\,z^2 \, \frac{\Lambda}{l^4} \Bigr ].
\end{multline}
All of these integrals converge, so we evaluate $X$ as:
\begin{multline}
  \label{X_gauss_eval}
  X = \rz \frac{2 \LE^2}{l^2}\\
  +\rz\, \frac{\LE \sqrt{\pi}}{4 l^3 \Lambda^2}\,  \Biggl \lbrace
  \Bigl (l^4 - 2 l^2 \Lambda\LE + 4 \Lambda^2 \LE^2 \Bigr ) \exp \left [
    \frac{l^2}{4} \, \left( \frac{1}{ \Lambda} + \frac{1}{ \LE} \right
      )^2 \right ] \erfc \left [ \frac{l}{2} \left(
      \frac{1}{\Lambda} + \frac{1}{ \LE} \right ) \right ] \\
    - \Bigl (l^4 + 2 l^2 \Lambda\LE + 4 \Lambda^2 \LE^2 \Bigr ) \exp \left
    [ \frac{l^2}{4} \, \left( \frac{1}{\Lambda} - \frac{1}{\LE} \right
      )^2 \right ] \erfc \left [ \frac{l}{2} \left(
      \frac{1}{\Lambda} - \frac{1}{\LE} \right ) \right ] \Biggr
    \rbrace.
\end{multline}

After inserting Eq. \ref{X_gauss_eval} into
Eq. \ref{eval_ave_eqn_LAM}, we solve for \LE\ over three decades in
the correlation length, $l$, and in the disorder variance, $\rz$
(Fig. \ref{lambdaEff}). At large correlation length the effective
penetration depth is larger than the average value, representing
\textit{ suppressed } Meissner screening. Conversely, at small
correlation length the effective penetration depth is smaller than the
average, indicating \textit{ enhanced } screening. The separatrix,
where $\LE = \Lambda$ for all values of $\rz$, occurs near $ l = 1.643
\Lambda$. Note that the system is not symmetric about the separatrix,
although it becomes more symmetric as $\rz \rightarrow 1$. This is
true for both linear and logarithmic spacing around the separatrix. In
other words, neither $|\LE(l_s + \Delta l) - \ave{\lambda}| = |\LE(l_s
- \Delta l) - \ave{\lambda}|$ nor $|\LE(a l_s) - \ave{\lambda}| =
|\LE(l_s/a) - \ave{\lambda}|$ are true, where $l_s$ denotes the
separatrix, and $a$ is an arbitrary positive real number. As expected,
$\LE \rightarrow \Lambda$ as $\rz \rightarrow 0.$ Yet even at small
disorder, \LE\ has variations on the one percent scale, shown by the
contours in Fig. \ref{lambdaEff}. As we will discuss below,
sub-percent variations of \LE\ could be significant in the context of
a typical measurement of $\Delta\lambda(T)$.

\begin{figure}[p]
  \includegraphics[width = 0.8\columnwidth]{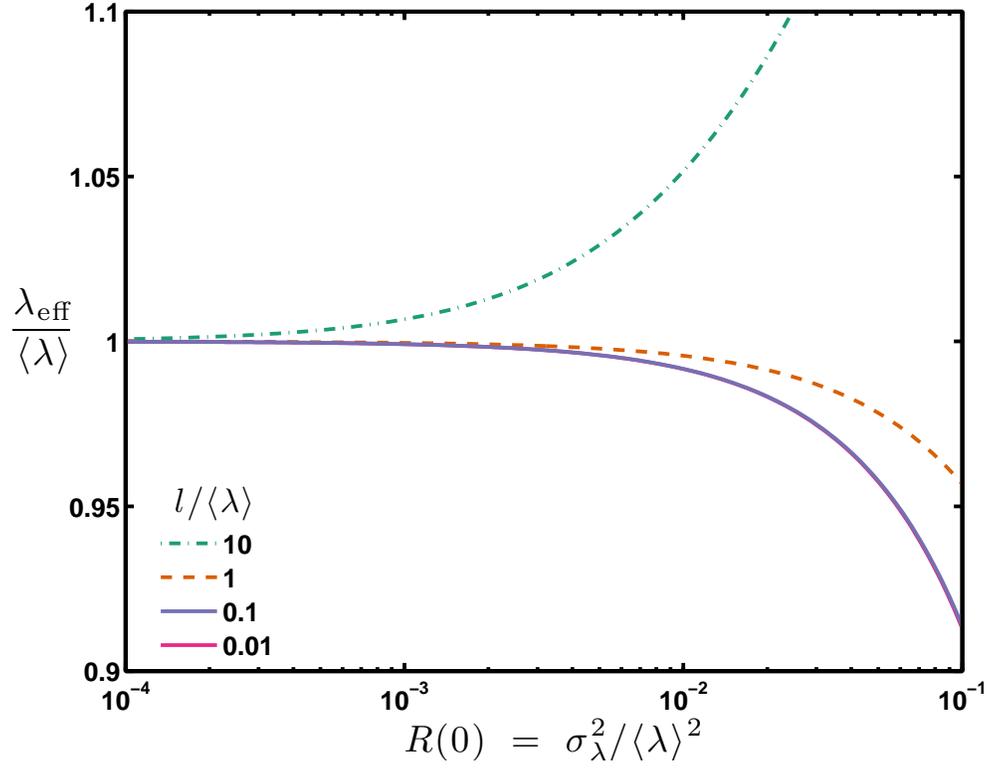}
  \caption{\label{lineCuts} The screening can either be enhanced ($\LE
    < \ave{\lambda}$) or suppressed ($\LE > \ave{\lambda}$), depending
    on the correlation length. The curves for $l = 0.1 \ave{\lambda}$
    and $l = 0.01\ave{\lambda}$ overlap. }
\end{figure}

The trends in \LE\ can also be seen in Fig. \ref{lineCuts}, where we
plot $\LE/\Lambda$ vs. $\rz$ at fixed correlation length. All
three curves taper to $\LE = \Lambda$ as the magnitude of disorder
decreases. At large correlation length, in this case $l = 10 \Lambda$,
\LE\ increases by ten percent when $\rz = 0.02$. The effect at
small correlation is more modest, but still reaches nearly ten percent by
the time $\rz = 0.1$ when $l = 0.1 \Lambda$.

The penetration depth has a temperature dependence that it inherits
from the underlying disordered superconducting state. It is natural to
expect that \rz\ and $l$ will have a temperature dependence of their
own, which will create a temperature-induced change in \LE. This
change contributes to any measurement of $\lambda(T)$, but is not
related to the gap structure in momentum space, because it arises from
the spatial arrangement of the superconducting state. If we neglected the
spatial variation of $\lambda$ we would erroneously attribute the
entire temperature dependence to the order parameter.

\subsection{General Mat\'{e}rn Correlations}
\label{sec:gener-matern-corr}
To understand the impact of the smoothness of $\lambda(\vecr)$ on the
measured penetration depth, \LE, we now consider the general case of
Mat\'{e}rn covariance.  Recall that the parameter $\nu$ controls the
smoothness of the penetration depth. With the correlation function
defined by Eq. \ref{matern}, we evaluate Eq. \ref{X_gped_sincos}:
\begin{multline}
  \label{X_insert_matern_Gamma}
  X = -\frac{\rz}{2^{\nu -1} \Gamma(\nu)} \int_0^\infty \!
  \mathrm{d}z\ e^{-z/\Lambda}\, \mathrm{sinh}\negmedspace
  \left(\frac{z}{\LE} \right )\ \frac{\LE}{\Lambda^2 l^4}\, \Biggl [
    l^4 \omegy^\nu K_\nu\negmedspace \omegy \\ + 4 \nu l^2 \left(
    \Lambda^2 + \Lambda z + \LE^2 \right) \omegy^{\nu-1}
    K_{\nu-1}\negmedspace \omegy \\ + 16 \nu^2 \LE^2 \Lambda ( \Lambda
    + z ) \omegy^{\nu-2} K_{\nu-2}\negmedspace \omegy \Biggr ] \\
  + \frac{\rz}{2^{\nu -1} \Gamma(\nu)} \int_0^\infty \!
  \mathrm{d}z\ e^{-z/\Lambda}\, \mathrm{cosh}\negmedspace
  \left(\frac{z}{\LE} \right ) \frac{4 \nu \LE^2}{\Lambda^2 l^4 }
  \Biggl [ l^2 z \omegy^{\nu-1}K_{\nu-1}\negmedspace\omegy \\ + 4 \nu
    \Lambda ( \Lambda + z) z \omegy^{\nu-2}
    K_{\nu-2}\negmedspace\omegy \Biggr].
\end{multline}
These integrals can be evaluated using equation 6.621.3 in
Gradshteyn and Ryzhik:\cite{gradshteyn_table_1980}
\begin{multline}
  \label{GR}
  \int_0^\infty \! x^{\mu-1}e^{-\alpha x} K_\nu (\beta x)
  \mathrm{d}x = \\ \sqrt{\pi} \frac{(2 \beta)^\nu}{(\alpha +
    \beta)^{\mu+\nu}}\, \frac{\Gamma(\mu+\nu)
    \Gamma(\mu-\nu)}{\Gamma(\mu+\tfrac{1}{2})} \, _2F_1 \left( \mu + \nu ,
  \nu + \tfrac{1}{2} ; \mu + \tfrac{1}{2} ;
  \frac{\alpha-\beta}{\alpha+\beta} \right ),
\end{multline}
which requires $\mathrm{Re}\, \mu > |\mathrm{Re}\, \nu |$ and $
\mathrm{Re} \, (\alpha + \beta ) > 0$. The function $_2 F_1(a,b;c;z)$
is Gauss' hypergeometric function. Using the integral in Eq. \ref{GR}
to evaluate Eq. \ref{X_insert_matern_Gamma}, we find the constraints
\begin{equation}
  \nu > \frac{3}{2} \quad \mathrm{and} \quad \frac{\LE}{\Lambda} >
  \frac{l}{l+2 \Lambda \sqrt{\nu}}.
\end{equation}

The full solution for $X$ is then:
\begin{multline}
  \label{X_matern_eval}
  X = \frac{\rz \,\sqrt{\pi}}{\Gamma(\nu)} \left(\frac{4 \nu}{l^2}
  \right)^\nu \frac{\LE}{\Lambda} \Biggl \lbrace
  \frac{\LE^2 + \Lambda^2}{2 \Lambda} \frac{\Gamma(2\nu-1)}{\Gamma(\nu
    + \tfrac{1}{2})} \Bigl [ a^{-(2\nu-1)} F1(\square) - b^{-(2\nu-1)}
    F1(\clubsuit) \Bigr ] \\
  + \frac{\Gamma(2\nu)}{\Gamma(\nu+\tfrac{3}{2})} \left [ \frac{\LE +
      \Lambda}{2 \Lambda} a^{-2\nu} F2(\square) + \frac{\LE -
      \Lambda}{2 \Lambda} b^{-2\nu} F2(\clubsuit) \right ]\\
  + \frac{1}{\Lambda} \frac{\Gamma(2\nu+1)}{\Gamma(\nu+\tfrac{3}{2}) }
  \Bigl [ a^{-(2\nu+1)} F3(\square) - b^{-(2\nu+1)} F3(\clubsuit)
    \Bigr] \\
  + \frac{\LE^2 \Lambda}{4} \frac{\Gamma(2\nu-3)}{\Gamma(\nu-1)}
  \Bigl[ a^{-(2\nu-3)} F4(\square) - b^{-(2\nu-3)} F4(\clubsuit) \Bigr
  ] \\
  + \frac{\LE}{4} \frac{\Gamma(2\nu-2)}{\Gamma(\nu + \tfrac{1}{2})}
  \Bigl [ \bigl( \LE + \Lambda \bigr ) a^{-(2\nu-2)} F5(\square) -
    \bigl ( \LE - \Lambda \bigr ) b^{-(2\nu-2)} F5(\clubsuit) \Bigr ] \\
  + \frac{\LE}{2} \frac{\Gamma(2\nu-1)}{\Gamma(\nu + \tfrac{3}{2})}
  \Bigl [ a^{-(2\nu-1)} F6(\square) + b^{-(2\nu-1)} F6(\clubsuit)
    \Bigr] \Biggr \rbrace,
\end{multline}
where we have introduced the variables
\begin{eqnarray*}
  a &=& \frac{1}{\Lambda} + \frac{1}{\LE} + \frac{2\sqrt{\nu}}{l}, \\ b
  &=& \frac{1}{\Lambda} - \frac{1}{\LE} + \frac{2\sqrt{\nu}}{l},
  \\ \square &=& \frac{l(\LE + \Lambda) - 2\LE \Lambda
    \sqrt{\nu}}{l(\LE + \Lambda) + 2\LE \Lambda \sqrt{\nu}},
  \\ \clubsuit &=& \frac{l(\Lambda - \LE) + 2\LE \Lambda
    \sqrt{\nu}}{l(\Lambda - \LE) - 2\LE \Lambda \sqrt{\nu}},
\end{eqnarray*}
and functions
\begin{eqnarray*}
  F1(\cdot) &=&  _2F_1(2\nu-1,\nu-\tfrac{1}{2};\nu+\tfrac{1}{2};\cdot),\\
  F2(\cdot) &=&  _2F_1(2\nu,\nu-\tfrac{1}{2};\nu+\tfrac{3}{2};\cdot),\\
  F3(\cdot) &=&  _2F_1(2\nu+1,\nu+\tfrac{1}{2};\nu+\tfrac{3}{2};\cdot),\\
  F4(\cdot) &=&  _2F_1(2\nu-3,\nu-\tfrac{3}{2};\nu-\tfrac{1}{2};\cdot),\\
  F5(\cdot) &=&
  _2F_1(2\nu-2,\nu-\tfrac{3}{2};\nu+\tfrac{1}{2};\cdot),\quad \text{and}\\
  F6(\cdot) &=&  _2F_1(2\nu-1,\nu-\tfrac{3}{2};\nu+\tfrac{3}{2};\cdot).\\
\end{eqnarray*}

Inserting this expression for $X$ into Eq. \ref{eval_ave_eqn_LAM}, we
can solve for \LE\ after choosing a value for the smoothness parameter
$\nu$. In Fig. \ref{lambdaMatern}, we have chosen $\nu = 2$, close to
the lower bound of $\tfrac{3}{2}$ required for convergence of $X$. The
results are almost identical to the case of squared exponential
correlations (Fig. \ref{lambdaEff}); evidently \LE\ is not
much affected by changes in the smoothness of $\lambda(\vecr)$ for the
Mat\'{e}rn family of correlation functions. The qualitative features
of interest to us are still present: there are regions of enhanced
screening and regions of suppressed screening, the effect grows on
increasing the variance of $\lambda(\vecr)$, and changes in \LE\ at
the one percent level persist down to small disorder. Quantitatively,
the results in Figures \ref{lambdaMatern} and \ref{lambdaEff} differ
by five percent in the region near $l = 1$ and $\rz = 1$, where the
difference is largest.

\section{Discussion}

The measured $\lambda(T)$ in a non-uniform superconductor will be
determined by both the momentum space gap structure and the real space
variations of the penetration depth. We calculate the influence of
spatial fluctuations in the penetration depth by solving the
stochastic London equation in the limit of small fluctuations. This
gives an equation (Eq. \ref{eval_ave_eqn_LAM}) for the
disorder-averaged magnetic field in terms of the penetration depth
correlation function. We then solve this equation for two example
correlation functions to find \LE, the decay length of the
disorder-averaged field, which we identify as the penetration depth
measured experimentally. We find that \LE\ can be either smaller or
larger than the average penetration depth, depending on the
correlation length of $\lambda$. More importantly, the variance and
correlation length of $\lambda$ will likely change with temperature,
endowing the experimentally measured penetration depth with
temperature dependence that is unrelated to the superconducting order
parameter.

This work shows that there can be a disorder-induced change of the
penetration depth that is not caused by the structure of the
superconducting gap in momentum space. Rather, it reflects the real
space variations of the order parameter. An interpretation that
assumed a spatially uniform penetration depth would infer a larger
modulation of $\Delta(\vk)$ than truly exists. Because $\Delta(\vk)$
is the starting point for investigations of the mechanism of the
superconductor, this omission could lead us astray when we seek to
determine the underlying mechanism.

How significant is the effect of disorder-induced change in the
penetration depth, given that $\LE/\ave{\lambda}$ approaches 1 over
a large segment of the \rz-$l$ plane? Modern measurements can
routinely resolve sub-nanometer changes in the penetration depth;
\cite{hardy_precision_1993,prozorov_magnetic_2006,luan_local_2011} in
cuprates and pnictides the penetration depth is approximately 200 nm,
and a 1-nm change in $\lambda$ yields $\tfrac{\Delta\lambda}{\lambda}$
of 0.5\% -- making even small changes in $\LE/\ave{\lambda}$
potentially significant.

Two issues are worth emphasizing. First, we have made no assumption
about the distribution of $\lambda(\vecr)$, i.e., whether it is
normally distributed or follows a different probability
distribution. However, the calculation presented here only extends to
second order, and any non-normality only enters at third order and
above. Second, \LE\ has a complicated dependence on the correlation
function \rarg, and we know neither its functional form nor its
temperature dependence. Hence we cannot make any tidy prediction for
the low-temperature behavior of $\lambda(T)$; there is no power-law to
be had.

\begin{figure}[p]
  \includegraphics[width = 0.8\columnwidth]{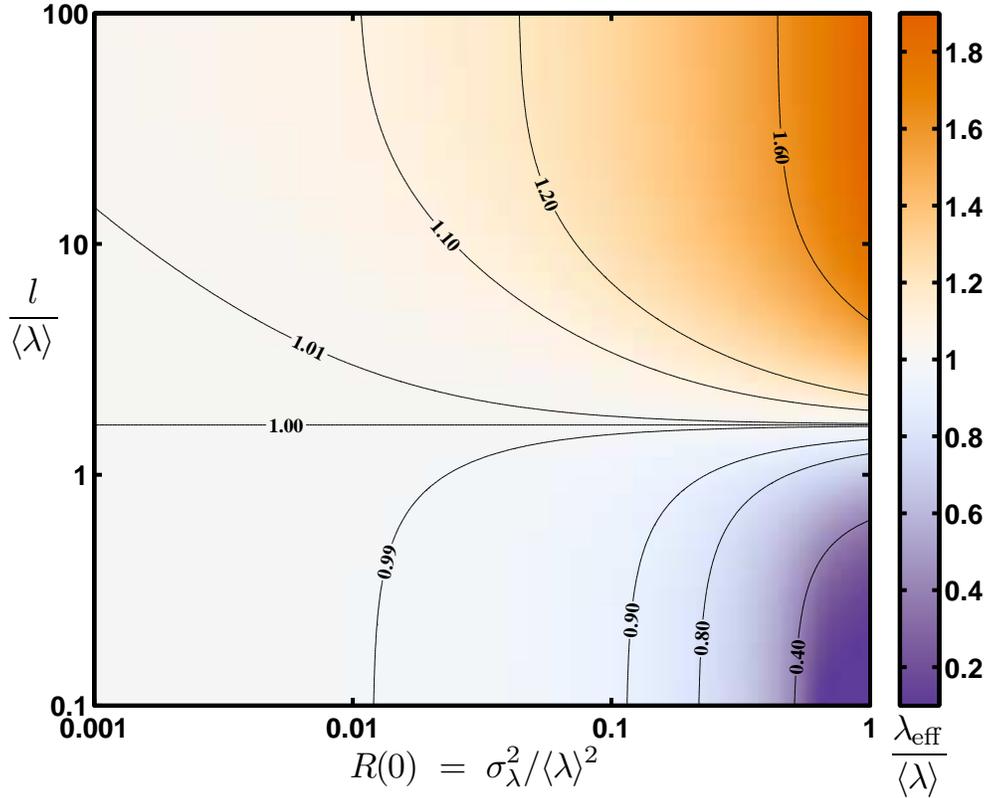}
  \caption{\label{lambdaEff} The effective penetration depth is a
    strong function of the parameters that characterize the
    distribution of local penetration depths. Here we show the value
    of \LE\ as the correlation length, $l$, and variance, $\rz$, run
    across three orders of magnitude. This figure considers the case
    of squared exponential correlations in the penetration depth; a
    different case is shown in Fig. \ref{lambdaMatern}. The most
    important features of this color plot are the range of
    $\LE/\ave{\lambda}$ and the appearance of values both above and
    below 1. The calculation is valid when $\rz \ll 1$, but we show
    the region with $\rz > 0.1$ to emphasize the trends seen.  Any
    temperature dependence in $l$ or $\rz$ will contribute to
    $\lambda(T)$. This temperature dependence is not accounted for by
    the superconducting gap.  }
\end{figure}

\begin{figure}[p]
  \includegraphics[width = 0.8\columnwidth]{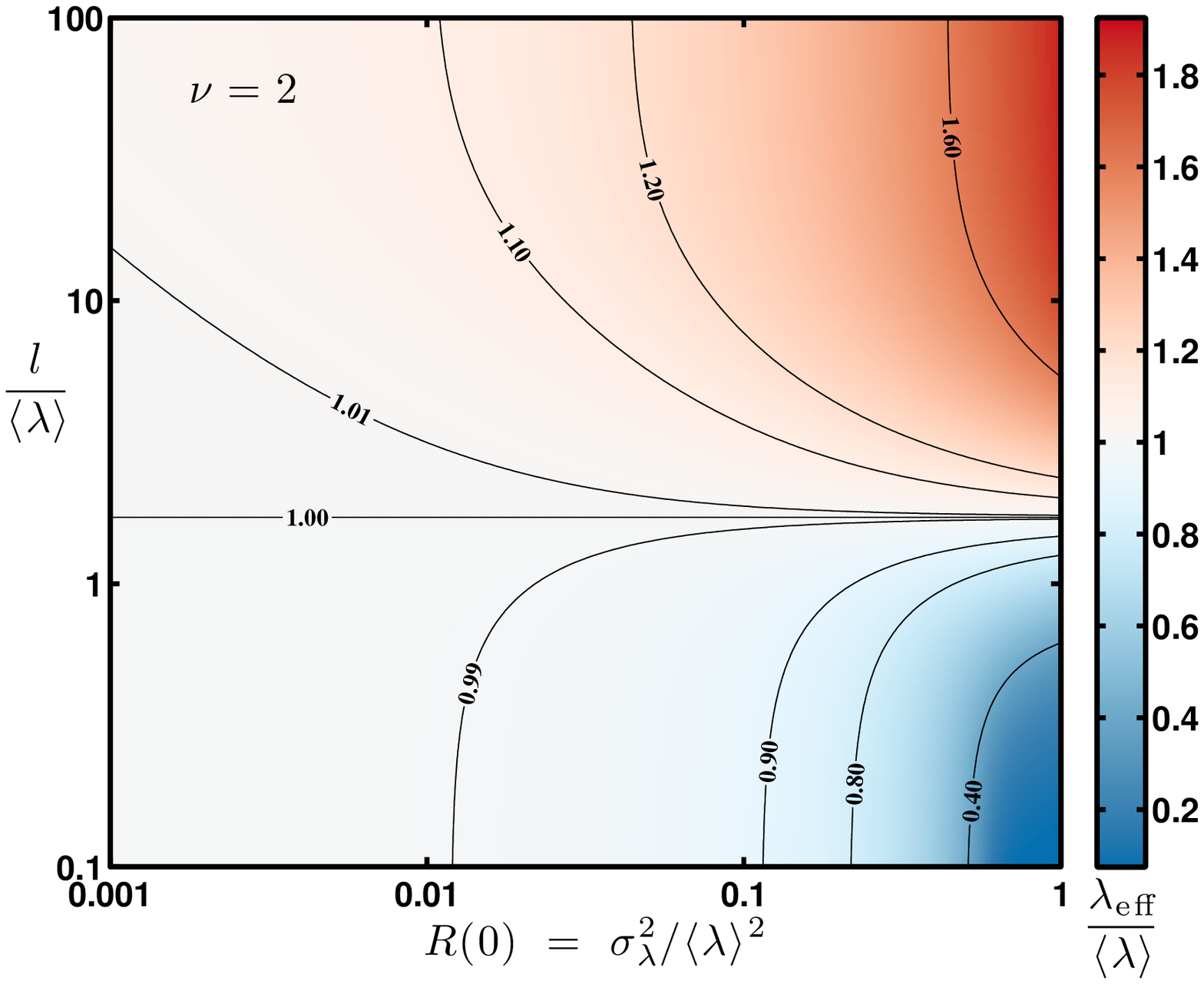}
  \caption{\label{lambdaMatern} The effective penetration depth for
    Mat\'{e}rn correlations when $\nu = 2$ (shown here) has strong
    similarities to Fig. \ref{lambdaEff}, which represents the
    limiting case where $\nu \rightarrow \infty$. These similarities
    imply that the smoothness of the random penetration depth does not
    strongly affect \LE.}
\end{figure}

Even without perfect knowledge of \rarg, it may be possible learn more
about \LE\ by taking advantage of the general constraints that apply
to all correlation functions.
\cite{matern_spatial_1986,cressie_statistics_1991} In particular, the
strong similarities between the two cases presented here (Figs.
\ref{lambdaEff} and \ref{lambdaMatern}) lead us to expect
qualitatively similar behavior in \LE\ for most possible correlation
functions.

To make a stronger statement about $\lambda(T)$, we need to determine
the local superconducting properties of a given chemically doped and
intrinsically disordered material, which naturally depends on the
microscopic details of the superconducting mechanism. Although it
should be possible to extract a local penetration depth or superfluid
density from numerical methods such as solving the
Bogoliubov~\!\!-~\!\!de Gennes equations on a lattice, to the best of
our knowledge this has never been attempted. Several groups have
calculated the \textit{disorder-averaged} superfluid stiffness using
this approach, for both s-wave\cite{ghosal_inhomogeneous_2001} and
d-wave\cite{franz_critical_1997,ghosal_spatial_2000} models. The full
temperature dependence of the disorder-averaged superfluid density can
also be calculated,\cite{Das:11055109:2011} but is incomplete,
for we have shown that the real space inhomogeneity of the
superconducting state also contributes to the temperature dependence.

The larger message is that some measured properties of disordered
superconductors will not be determined by their disorder averages
alone; inhomogeneities can affect the measured properties in an
experiment-dependent manner. For example, the heat capacity will be
given by the disorder average because it is additive, but we have seen
that the penetration depth is non-trivially affected by the
disorder. Nonetheless, these two experiments are both traditionally
interpreted as measuring the same thing -- the magnitude of the
single-particle gap, $\Delta(\vk)$.

These results give a specific example of the potential impact of
spatial variation on measurements of the penetration depth.  With a
full consideration of the impact of spatial variation on different
measured quantities, as well as a complete understanding of how random
chemical doping gives rise to a non-uniform superconducting state, we
will be able to integrate a complete account of the effects of
disorder into our understanding of unconventional superconductivity.

\begin{acknowledgments}
We thank John Kirtley, Steve Kivelson, Jim Sethna, and J\"{o}rg
Schmalian for helpful discussions. We would also like to thank John
Kirtley for checking some of these calculations. This work
is supported by the Department of Energy, Office of Basic Energy
Sciences, Division of Materials Sciences and Engineering, under
contract DE-AC02-76SF00515
\end{acknowledgments}

\begin{thebibliography}{28}
\bibitem {london_footnote} The familiar relation $\vect{J} =
  -\vect{A}/\lambda^2$ is not valid when the order parameter is
  non-uniform.  Instead, we start with the Ginzburg-Landau equations,
  which can describe a non-uniform superconducting state. From the
  second Ginzburg-Landau equation we derive our starting point,
  Equation \ref{london_def}, a generalized London equation.
\end{thebibliography}
\end{document}